\documentclass[aps,prd,preprint,showpacs,nofootinbib,preprintnumbers]{revtex4}

\usepackage{epsfig}
\usepackage{amsmath}
\usepackage{amssymb}
\usepackage[dvipdfm]{hyperref} 
\def\sn{\,{\rm sn}}
\def\cn{\,{\rm cn}}
\def\dn{\,{\rm dn}}
\def\e{{\,\rm e}\,}
\def\d{{\rm d}}
\def\i{{\rm i}}
\def\K{K}

\def\D{{\cal D}}

\newcommand{\rf}[1]{(\ref{#1})}
\newcommand{\eq}[1]{Eq.~(\ref{#1})}
\renewcommand{\star}{*}
\def\be{\begin{equation}}
\def\ee{\end{equation}}
\def\beq{\begin{equation}}
\def\eeq{\end{equation}}
\def\bea{\begin{eqnarray}}
\def\eea{\end{eqnarray}}

\newcommand{\non}{\nonumber \\*}

\newcommand{\pintii}{\int_{-\infty}^{+\infty}\hspace{-2.45em}\not\hspace{2.0em}}
\newcommand{\pint}{\int}
\def\la{\lesssim}
\def\ga{\gtrsim}
\renewcommand{\Re}{{\rm Re}\,}
\renewcommand{\Im}{{\rm Im}\,} 

\begin{document}

\preprint{ITEP--TH--03/10}

\title{Quantum corrections from a path integral over reparametrizations}

\author{Yuri Makeenko}
\altaffiliation[Also at]{ the Institute for Advanced Cycling,
Blegdamsvej 19, 2100 Copenhagen \O, Denmark}

\affiliation{Institute of Theoretical and Experimental Physics\\
B.~Cheremushkinskaya 25, 117218 Moscow, Russia}
\email{makeenko@itep.ru}

\author{Poul Olesen$^*$} 
\affiliation{The Niels Bohr International Academy,
The Niels Bohr Institute\\
Blegdamsvej 17, 2100 Copenhagen \O, Denmark}
\email{polesen@nbi.dk}

\date{\today}

\begin{abstract}
We study the path integral over reparametrizations  
that has been proposed as an ansatz for the Wilson loops in 
the large-$N$ QCD and reproduces the area law in the classical limit
of large loops.
We show that a semiclassical expansion for a rectangular loop  captures 
the L\"uscher term associated with $d=26$ dimensions 
and propose a modification of the ansatz  which reproduces
the L\"uscher term in other dimensions, which is observed  in 
lattice QCD. 
We repeat the calculation for an outstretched ellipse 
advocating the emergence of an analog of the L\"uscher term 
and verify this result by a direct computation of the determinant 
of the Laplace operator and the conformal anomaly. 

\end{abstract}


\maketitle

\section{Introduction}

Quantum fluctuations of surfaces are important in many physical phenomena. It is
less known these fluctuations can sometimes be expressed in term of a
functional integral over reparametrizations of variables relevant in the
Feynman path integral. In this paper we shall consider the expression
\begin{equation}
W(C)\equiv \int {\cal D}\theta\,\e^{-KA[x(\theta)]},
\label{ansatz}
\end{equation}
where $x(\theta)$ is the boundary curve $C$ and where
\begin{equation}
A[x(\theta)]=\frac{1}{8\pi}~\int_0^{2\pi}\d\sigma \int_0^{2\pi}\d\sigma' ~
\frac{\left(x(\theta(\sigma))-x(\theta(\sigma'))\right)^2}
{1-\cos(\sigma-\sigma')}.
\label{D}
\end{equation}
The functional (\ref{D}) is known in mathematics as the Douglas integral 
\cite{Dou31}, whose minimum with respect to variations of the 
reparametrizations $\theta$ gives the minimal area
\begin{equation}
A\left[x(\theta_\star)\right]=S_{\rm min} (C) .
\label{p3}
\end{equation}
Here $\theta_\star(\sigma)$ is the saddle point of the integral 
(\ref{ansatz}). 
The functional integration (\ref{ansatz}) 
thus gives the area law to leading order.

The path integral over reparametrizations (\ref{ansatz}) was introduced
in this context in Ref.~\cite{cohen} in connection with an off-shell string 
propagator. More recently it was proposed by Polyakov \cite{Pol97} as
an ansatz for the Wilson loop in large $N$ QCD. The leading behavior obviously
gives the leading behavior of the Wilson loop, found in most string models,
where the bulk field $X^\mu(\tau,\sigma)$ satisfies the Dirichlet 
boundary condition
\begin{equation}
X^\mu(\tau,\sigma)|_{\rm boundary}=x^\mu (\sigma).
\label{p4}
\end{equation}
To derive Eq.~(\ref{p3}) from the Nambu-Goto action one can follow
Douglas \cite{Dou31}, or the more recent elegant paper by Migdal
\cite{Mig94}. 

The functional integral (\ref{ansatz}) can be expanded around the
saddle point $\theta_\star$. With $\theta(\sigma)=\theta_\star(\sigma)+
\beta (\sigma)$ we obtain
the first non-vanishing contribution
\begin{equation}
A_2=\frac{K}{8\pi}~\int_0^{2\pi}\d\theta\int_0^{2\pi}\d\theta'\,{\dot x}
(\theta)\cdot{\dot x}(\theta')\frac{\left(\beta(\theta)-\beta(\theta')\right)^2}
{1-\cos (\sigma_\star (\theta)-\sigma_\star (\theta'))}
\label{p5}
\end{equation} 
Here $\sigma_\star (\theta)$ is the inverse function of $\theta_\star(\sigma)$.

Some dynamical consequences of Eq.~(\ref{p5}) have been discussed by Rychkov
\cite{Ryc02} and by the authors \cite{MO09}. 
However, it is perhaps fair to say
 that the physical meaning of the fluctuation integral (\ref{p5}) is not
so clear. 

In this paper we shall show that the leading part of the fluctuations of
large loops  from
(\ref{p5}) are {\it transverse fluctuations of the minimal surface} embedded
by the curve $x(\theta_\star)$. This we have shown for a rectangle and an 
ellipse, but we suspect the result to be more general. 

To explain our result we mention that in the
Nambu-Goto action transverse fluctuations add a contribution
\begin{equation}
-\frac{d-2}{2}~{\rm tr ~log}~(-\partial^2)=\frac{(d-2)\pi}{24}~\frac{T}{R}
\label{lut}
\end{equation}
to the area term
for a large $R\times T$ rectangle with $T\gg R$. These quantum fluctuations
(\ref{lut}) are called the L\"uscher term \cite{luscher}. By lattice 
Monte Carlo
calculations in three and four dimensions, this term has been found to occur
\cite{monte} in quenched $SU(N)$ for various $N'$s. Thus, for large distances 
the two leading terms from the Nambu-Goto action describe QCD quite 
well. Therefore, for a rectangular boundary curve the $T/R$ term
can be identified with transverse fluctuations of the {\it minimal surface}.

With these remarks in mind we now give our main result: {\it after
integration over the fluctuations $\beta$ in the functional integral }
\begin{equation}
\int {\cal D}\beta\e^{-A_2}
\label{p7}
\end{equation}
{\it we obtain as the leading contribution the L\"uscher term corresponding to
d=}26. In general, we obtain the $d$-dimensional contribution (plus the area 
term) from
\begin{equation}
\int {\cal D}\theta \e^{-KA[x(\theta)]}~(\det O)^{-(d-26)/48}.
\label{p8}
\end{equation} 
Here $O$ is the operator which emerges in the semiclassical expansion of 
$A\left[x(\theta)\right]$ to quadratic order as exhibited in Eq.~(\ref{p5}). 
This modification of Eq.~(\ref{ansatz})  does
not effect the classical limit leading to the area law, and has the meaning 
of a pre-exponential in the semiclassical approximation.
Thus, this equation gives the leading effective QCD string behavior in terms 
of functional integration over reparametrizations.

We mention that our considerations may have potential applications
in condensed matter. For example, in the inverse square XY model
one encounters  expressions somewhat similar to Eq.~(\ref{p5}), see
for example Ref.~\cite{hof}. We shall, 
however, not pursue this track in the present paper.

The plan of this paper is the following: in Sect.~\ref{s:2} 
we discuss the general framework for the semiclassical approximation, 
and in Sect.~\ref{s:3} we carry out the
functional integral over reparametrizations for a rectangle. A similar 
calculation for an ellipse is done in Sect.~\ref{s:4}. 
In Sect.~\ref{s:5} we derive the generalization given by Eq.~(\ref{p8}) 
and in Sect.~\ref{s:6} we make some
conclusions. A number of more technical points
have been discussed in some Appendices.

\section{A semiclassical correction\label{s:2}}

The path integral over reparametrizations~\rf{ansatz} reproduces the 
exponential of the minimal area in the classical limit 
$\K S_{\rm min}\to\infty$.
To calculate the semiclassical correction, 
we expand
\be
\theta(\sigma)=\theta_*(\sigma)+\beta(\sigma)
\label{expa}
\ee
or
\be
\sigma(\theta)=\sigma_*(\theta)+\beta(\theta) \,,
\label{expa1}
\ee
where
\be
\beta(0)=\beta(2\pi)=0
\label{bbcb}
\ee
and expand the Douglas integral to
quadratic order in $\beta$ around the classical trajectory $\theta_*(\sigma)$. 
This expansion makes sense because typical trajectories in the path
integral over reparametrizations~\rf{ansatz} are smooth as 
$\K S_{\rm min}\to\infty$ and have Hausdorff dimension one~\cite{BM09}.

Substituting~\rf{expa} into Douglas' integral~\rf{D} and expanding in $\beta$,
we find that the linear term vanishes 
because $\theta_*(\sigma)$ is the minimum, while the quadratic part
reads
\be
A_2[\beta(\theta)]
=\frac{K}{8\pi}\int_0^{2\pi}\d\theta_1\int_0^{2\pi}\d\theta_2\,{\dot x}
(\theta_1)\cdot{\dot x}(\theta_2)\,\frac{\left(\beta(\theta_1)-\beta(\theta_2)
\right)^2}
{1-\cos (\sigma_\star (\theta_1)-\sigma_\star (\theta_2))}\,.
\label{S2}
\ee
The function $\beta(\theta)$  obeys
\be
\dot \beta (\theta) \geq -\dot\sigma_*(\theta)
\label{restriction}
\ee
for the derivative of the reparametrizing function to be positive.
This is always satisfied if $\beta$ is small and smooth enough.

In order to calculate the semiclassical correction to the area law, we need
to do the Gaussian integral
\be
\int \D\beta(\theta) \e^{-A_2[\beta(\theta)]}
\label{II2}
\ee
with $A_2[\beta(\theta)]$ given by \eq{S2}.

The typical values of $\beta$, which
are essential in the path integral over $\beta$ in \rf{II2},
are $\beta\sim 1/\sqrt{K S_{\rm min}}$, i.e. 
 small for $\sqrt{K S_{\rm min}}  \gg1$.
Hence, the higher terms of an expansion 
of $A[\theta_*(\sigma)+\beta]$ in
$\beta$ are suppressed~\cite{Ryc02} at large $\sqrt{K  S_{\rm min}} $.
The loop expansion goes in the parameter $1/K S_{\rm min}$ and only one loop
contributes with the given accuracy. 

A comment is needed about the measure for the integration over $\beta(\theta)$.
As is explained in Appendix~\ref{appA}, the measure for the integration
over $\sigma(\theta)$ involves a factor
\be
\frac{1}{\sigma'}=\frac{1}{\sigma'_*+\beta'}=
\frac{1}{\sigma'_*}-\frac{\beta'}{(\sigma'_*)^2}+
\frac{(\beta')^2}{(\sigma'_*)^3}+
{\cal O}\left((\beta')^3\right).
\ee
Because  $\beta\sim 1/\sqrt{K S_{\rm min}}$, the second and third terms
on the right-hand side are not essential to the given order.
Therefore, the measure for the path integration over $\beta(\theta)$
in \eq{II2} is the usual one for smooth functions $\beta(\theta)$, while
this factor will show up to the next order in $ 1/{K S_{\rm min}}$.

\section{Path integral over reparametrizations: rectangle\label{s:3}}

In this section we show how the path integral over 
reparametrizations captures the L\"{u}scher term for a rectangle.

The conformal map of the upper half plane onto the interior of
a rectangle is given by  the Schwarz--Christoffel mapping
\be
\omega =A F\left(\frac{z}{\sqrt{\mu}},\mu\right)
-\i \frac{AK\left(\sqrt{1-\mu^2}\right)}2,
\ee
where 
\be
F\left(\frac{z}{\sqrt{\mu}},\mu\right)=\int_0^z 
\frac{\d x}{\sqrt{\mu-x^2}\sqrt{1-\mu x^2}}
\ee 
is the incomplete elliptic integral of the first kind.
The two parameters $A$ and $\mu$ are related to the coordinates
of the vertices of the rectangle by
\be
A K(\mu)=\frac R2,\qquad A K\left(\sqrt{1-\mu^2}\right)=T,
\label{RandT}
\ee
where $K(\mu)=F(1,\mu)$ is the complete elliptic integral of the first kind.
In deriving \eq{RandT} we used the important identity 
\be
F(1/\mu,\mu)=K(\mu)+\i K\left(\sqrt{1-\mu^2}\right).
\ee
Equations~\rf{RandT} relates $\mu$ to the ration of $R/T$ as
\be
\frac{2T}{R}=\frac{ K\left(\sqrt{1-\mu^2}\right)}{K(\mu)}.
\label{muvsT/R}
\ee
In the limit $T/R\to\infty$, when $\mu\to0$, this equation simplifies to
\be
\pi \frac{T}R=\ln \frac4{\mu}\, .
\label{muto0}
\ee

When $z=s$ runs along the real axis, the variable $\omega$ runs along
the boundary of the rectangle with $s=-1/\sqrt{\mu}, -\sqrt{\mu}, 
+\sqrt{\mu}, +1/\sqrt{\mu}$ ($\mu<1$)
mapped, respectively, onto the vertices of the rectangle: 
$(-R/2,T/2),(-R/2,-T/2),(R/2,-T/2), (R/2,T/2)$. The given choice of the
argument of the mapping preserves the symmetry
$s\to1/s$.

When $z$ has positive imaginary part,
the coordinates
\be
X_1(z)=A\, \Re F\left(\frac{z}{\sqrt{\mu}},\mu\right),\qquad 
X_2(z)= A\,\Im F\left(\frac{z}{\sqrt{\mu}},\mu\right)-
\frac{AK\left(\sqrt{1-\mu^2}\right)}2
\label{KSmap}
\ee
take their values inside the rectangle. These coordinates are conformal.
For this reason we have
\be
x_1(t_*(s))=A \,\Re F\left(\frac{s}{\sqrt{\mu}},\mu\right),\qquad 
x_2(t_*(s))= A\,\Im F\left(\frac{s}{\sqrt{\mu}},\mu\right)-
\frac{AK\left(\sqrt{1-\mu^2}\right)}2,
\label{Crecta}
\ee
whose implementation for the function $t_*(s)$ is discussed below. 

The boundary contour given by \eq{Crecta} satisfies Douglas' 
minimization (see Appendix~\ref{appB}, \eq{Dou1}).
Correspondingly, the Douglas integral   
\be
\frac 1{4\pi}
\int\nolimits_{-\infty}^{+\infty} \d s 
\int\nolimits_{-\infty}^{+\infty} \d s' \,\frac{ \left[ 
 x(t_*(s_1))- x(t_*(s_2))\right]^2}{(s-s')^2} = 
2 A^2 K(\mu) K\left(\sqrt{1-\mu^2}\right) =R T
\label{71}
\ee 
as it should. We have verified these two equations numerically.

A natural parametrization of the boundary of a rectangle is through
$\tau\in S^1$:
\begin{subequations}
\bea
x_1= \frac T2 \tan \tau \,,~~~ x_2= -\frac T2\qquad&& 
- \arctan \frac R T\leq\tau\leq \arctan \frac R T \\*
x_1= \frac R2 \,,~~~ x_2= -\frac R2 \cot \tau \qquad&& 
 \arctan \frac R T \leq\tau\leq \pi-\arctan \frac R T \\*
x_1= \frac T2 \tan \tau \,,~~~ x_2= \frac T2\qquad&& 
\pi- \arctan \frac R T\leq\tau <\pi
\eea
\label{xtau}
\end{subequations}
and analogously for negative $\tau$. Introducing 
\be
t=\tan\frac{\tau}{2},
\ee
we rewrite \eq{xtau} as
\begin{subequations}
\bea
x_1=  T \frac{t}{1-t^2}\,,~~~ x_2= -\frac T2\qquad&& 
-  \frac  {\sqrt{T^2+R^2}-T}R\leq t\leq \frac  {\sqrt{T^2+R^2}-T}R \\*
x_1= \frac R2 \,,~~~ x_2=  R \frac{t^2-1}{4t} \qquad&& 
 \frac {\sqrt{T^2+R^2}-T}R\leq t\leq \frac  {\sqrt{T^2+R^2}+T}R  \\*
x_1=  T \frac{t}{t^2-1}\,,~~~ x_2= \frac T2\qquad&& 
 \frac  {\sqrt{T^2+R^2}+T}R\leq t <+\infty. 
\eea
\label{xt} 
\end{subequations}
\indent To relate $t$ to $s$, we identify
\begin{subequations}
\bea
\frac{R}{2K(\mu)} F\left(\frac{ s}{\sqrt{\mu}},\mu\right)
&=& T \frac{t}{1-t^2}\,,\label{I1}\\* \hbox{for}~~
-\sqrt{\mu} \leq s \leq \sqrt{\mu}~~&\hbox{or}&~~  
-  \frac  {\sqrt{T^2+R^2}-T}R\leq t\leq \frac  {\sqrt{T^2+R^2}-T}R
\non
\frac{R}{2K(\mu)} \int^{s}_{\sqrt{\mu}} 
\frac{\d x}{\sqrt{x^2-\mu}\sqrt{1-\mu x^2}}- \frac T2&=&  R \frac{t^2-1}{4t}
\label{I2}\\* 
 \hbox{for}~~\sqrt{\mu} \leq  s \leq 1/\sqrt{\mu}~~&\hbox{or}&~~
 \frac {\sqrt{T^2+R^2}-T}R\leq t\leq \frac  {T+\sqrt{T^2+R^2}}R
\,. \nonumber 
\eea
\label{st} 
\end{subequations}
\hspace*{-1mm}Solving the quadratic equation for $t$ versus $s$, 
we obtain the minimizing function $t_*(s)$, which obviously obeys
the boundary condition
\be
t(0)=0.
\label{bbc}
\ee

The symmetry $  s\to 1/  s$ plays apparently an important role.
It guarantees that the points $-\infty$, $-1$, $0$, $+1$, $+\infty$ 
are mapped onto themselves under the reparametrization 
$s\to t_*(s)$:  $t_*(-\infty)=-\infty$, $t_*(-1)=-1$, $t_*(0)=0$, 
$t_*(+1)=+1$, $t_*(+\infty)=+\infty$. 

It is convenient to invert \eq{st} using the Jacobi elliptic functions.
Inverting \eq{I1}, we get
\bea
{s}&=&{\sqrt{\mu}}\,
{\rm sn}\left(\frac{2K(\mu)T}{R}\,\frac{t}{1-t^2},\mu \right),
\label{3}
\\* \hbox{for}~~
-\sqrt{\mu} \leq s \leq \sqrt{\mu}&~~&\hbox{or}~~  
-  \frac  {\sqrt{T^2+R^2}-T}R\leq t\leq \frac  {\sqrt{T^2+R^2}-T}R \,.
\nonumber
\eea

The function sn has a nice trigonometric expansion in the parameter (nome)
(Ref.~\cite{GR}, 8.146.1)
\begin{equation}
\exp{\left(-\pi \frac{K(\sqrt{1-\nu^2})}{K(\nu)}\right)}\approx (\mu/4)^2\ll1,
\label{5}
\end{equation}
giving 
\be
  s\approx\sqrt{\mu} \sin \left(  \frac{\pi T}{R} \frac{t}{(1-t^2)}\right).
\label{tts1}
\ee
This formula is applicable 
for $-R/2T <t<+ R/2T$, when $-\sqrt{\mu}<s<+\sqrt{\mu}$.

We can proceed in the same way with \eq{I2}, whose inverse is
\begin{equation}
s=\sqrt{\mu}\,{\rm sn}\left(K(\mu)+\i K(\mu)\left(\frac{t^2-1}{2t}+
\frac{T}{R}\right),\mu\right).
\label{7}
\end{equation}
Using the addition formula (\cite{GR}, 8.156.1) 
and the reduction of ${\rm sn}(x,0)$ and
${\rm cn}(x,0)$ to $\sin x$ and $\cos x$, we obtain
\be
  s\approx \sqrt{\mu} \cosh \left[\frac\pi 4 \left(
t-t^{-1}+\frac{2T}R \right) \right].
\label{nsss}
\ee
However, this expansion is useless for large $t \to 2T/R$, 
due to the imaginary part of the argument of sn the expansion will 
involve hyperbolic functions with arguments that can be large.  
Instead we can use the expansion of sn in terms
of inverse sines (\cite{GR}, 8.147.1), where these sines can be large, so 
only the first term is relevant:
\begin{equation}
s\approx \frac 
1{\displaystyle{\sqrt{\mu}\,\sin \left(\frac{\pi}{2}+\i\frac{\pi}{4}\left(
\frac{t^2-1}{t}-\frac{2T}{R}\right)\right)}}=
 \frac{1}{\displaystyle{\sqrt{\mu} \cosh \left[\frac\pi 4 \left(
t-t^{-1}-\frac{2T}R \right) \right]}}\,.
\label{sss}
\ee
Equation~\rf{nsss} is applicable for $t\to R/2T$ from above and
\eq{sss} is applicable when $t\to 2T/R$.

The quadratic action, describing Gaussian fluctuations around $t_*(s)$,
is like \rf{S2} rewritten for the real-axis parametrization: 
\be
A_2\left[\beta(t)\right] = \frac K{4\pi}  \int\d t_1 \pint \d t_2 
\frac{\dot x(t_1)\cdot \dot x(t_2)}{(s_*(t_1)-s_*(t_2))^2 } 
\left[\beta(t_1)-\beta (t_2)\right]^2.
\label{r-aA2}
\ee 
The Gaussian approximations is justified for large $K RT$, when 
\be
\beta(t)\sim \frac1{\sqrt{K R T}}.
\ee 

To calculate the path integral over the quantum fluctuations 
around the minimizing function $t_*(s)$, we need a mode expansion of
the infinitesimal reparametrizing function $\beta(t)$.
To get rid of the projective symmetry, we keep fixed 3 points, e.g.\
$-1$, $0$, $+1$ or $-\sqrt{\mu}$, $0$, $\sqrt{\mu}$ fixed: 
$\beta(-1)=0$, $\beta(0)=0$, $\beta(+1)=0$ or
$\beta(-t_*(-\sqrt{\mu}))=0$, $\beta(0)=0$, $\beta(t_*(\sqrt{\mu}))=0$. 

For each segment from $t_i$ to $t_f$, we consider the mode
expansion 
\be
\beta(t)=\sum_n c_n \sin \left( \pi n \frac{t-t_i}{t_f-t_i} \right),
\ee
obeying the boundary condition $\beta(t_i)=\beta(t_f) =0$.
This set of sines forms a complete basis on the given interval.
Actually we shall need the mode expansion for 4 segments attached to
$t=t_*(\pm \sqrt{\mu})\approx\pm R/2T$ because the large contribution of
$1/\mu$ will appear in $A_2$ only for those.

The appearance of the large factor $1/\mu$ is seen already 
from Eqs.~\rf{3}, \rf{7} because $s_*$ in the denominator in \eq{r-aA2} is
proportional to $\sqrt{\mu}$.
We need, however, to show that this $1/\mu$ is multiplied by a factor
$\sim 1$.

Let us analyze the contribution to \rf{r-aA2} that comes 
from $-R/2T<t_1,t_2<+ R/2T$, i.e.\ from the bottom side of the rectangle.
Introducing the variable
\be
y=\frac{2T}R t \,, \qquad  -1< y <1
\ee
and using \eq{tts1}, we write the contribution of this domain to $A_2$ as
\be
\frac {\K T^2}{4\pi \mu}  \int_{-1}^1\d y_1 \int_{-1}^1 \d y_2\, 
\frac{\left(\beta (y_1)-\beta(y_2)\right)^2}{
\left(\sin(\pi y_1/2)-\sin(\pi y_2/2)\right)^2 } \propto \frac{T}{\mu R} C    
\label{AA2}
\ee 
with a positive constant $C$.
A similar contribution appears 
if $t_1-R/2T\sim  t_2-R/2T\sim R^2/(2T)^2$ as is prescribed by \eq{nsss}
for both $t_1>R/2T$ and $t_2>R/2T$. An analogous
contribution (with possibly some powers of $T/R\propto\ln (4/\mu)$) emerges 
also when  $t_1< R/2T$ and $ t_2> R/2T $ 
or  $t_1>R/2T$ and $ t_2< R/2T $. 

Therefore, the path integral over the quantum fluctuations 
around the minimizing function $t_*(s)$ gives, using the $\zeta$-function
regularization and \eq{muto0}
(and disregarding the logs to the order in consideration)
\be
\prod_{\rm modes} {\sqrt{\mu}} \propto \left(\frac1
 {\sqrt[4]{\mu}}\right)^4 \propto \exp{\left(\frac{\pi T}{R} \right)},
\ee
where the 4th power is due to the four sets of modes%
\footnote{This is like
in the computation of the static potential for the Polyakov string in
Ref.~\cite{JM93}.}.
This remarkably reproduces the L\"uscher term~\rf{lut}  
in $d=26$!

There was a subtlety in the derivation -- the appearance of a
logarithmic divergence at the corners of the rectangle
if $\dot \beta({\rm corners})\neq0$.
It is seen from \eq{AA2}, where the region near $y_1=y_2=1$ or  $y_1=y_2=-1$
(associated with  $t_1=t_2\to R/2T$ or $t_1=t_2\to -R/2T$ )
produces the logarithmic divergence
\be
\int_{-1+\delta}^{1-\delta}\d y_1 \int_{-1+\delta}^{1-\delta} \d y_2\, 
\frac{\dot\beta^2 (\pm1)}{
\left(y_1+y_2\mp2\right)^2 } =  \dot\beta^2 (\pm1)\ln \frac 1\delta 
\label{AAAA2}
\ee
with the upper (lower) sign referring to $y=1$ ($y=-1$).
The coefficient is nonvanishing  if $\dot \beta(\pm1)\neq0$. 
Analogously, the integral is logarithmically divergent at the corner, when 
$t_1,t_2> R/2T$ or $t_1< R/2T$, $t_2> R/2T$
and vise versa.

The logarithmic divergence can be regularized by smoothing 
the corners like in Refs.~\cite{luscher,DOP84}. 
It is clear from such a regularization that 
the contribution of trajectories with $\dot \beta(R/2T)\neq0$
to the path integral over $\beta(t)$ will be suppressed as the
smoothing is removed. Consequently, 
this corner divergence does not effect the
result of this section. In the next section we repeat the
consideration for the case of an ellipse, when there are no corners.

If $d\neq 26$, the asymptotic ansatz for the Wilson loops has
to be improved to get the correct factor $(d-2)/24$ 
in the L\"uscher term~\rf{lut}.
This issue will be described in Sect.~\ref{s:5}.

\section{Path integral over reparametrizations: ellipse\label{s:4}}

In this section we evaluate the path integral over 
reparametrizations for an ellipse and obtain a prediction
for the associated L\"{u}scher term.

The necessary formulas are given in Appendix~B of \cite{MO09} and are partially
reproduced in Appendix~B. 
We are interested in
the case of a very long ellipse when the ratio $b/a\to0$ and 
$\nu\to1$ according to 
\be
\ln \frac{a+b}{a-b}=\frac{\pi K(\sqrt{1-\nu^2})}{2K(\nu)}\,.
\label{goerlish}
\ee
Using the asymptote
\be
K(\nu)\stackrel{\nu\to1}\to \frac 12 \ln \frac 8{(1-\nu)},
\ee
we simplify \eq{goerlish} to
\be
\frac ba =\frac {\pi^2}{4\ln \frac 8{(1-\nu)}}.
\ee

For $\nu\to1$ the elliptic function simplifies and we have
\be
\theta_*(\sigma)=
\pi \left(\frac{\ln \frac{2\sigma+\sqrt{4\sigma^2+(1-\nu)^2}}{8}}
{\ln \frac8{(1-\nu)}}  +1\right)
\label{approxi}
\ee
for $-\pi/2<\sigma<\pi/2$.
Inverting \eq{approxi}, we find
\be
\sigma_*(\theta)=\frac{1-\nu}2 
\sinh\left(\frac{2\theta}{\pi}\ln\frac8{1-\nu}\right).
\label{inver}
\ee
This is quite similar to Eqs.~\rf{tts1} and \rf{nsss} for a rectangle  
with $\sqrt{\mu}$ replaced by $(1-\nu)$.

The calculation of the path integral over reparametrizations at one loop
is quite analogous to that for the rectangle described in the previous section.
We see that the large factor of $(1-\nu)^{-2}$ emerges 
 in \eq{S2} because $\sigma_*\propto (1-\nu)$ from \eq{inver}. 
To evaluate the coefficient, let us consider the domain of small 
$\theta_1$ and $\theta_2$: 
\be
\theta_1,\theta_2 \ll  \left(\ln \frac{8}{1-\nu}\right)^{-1}\,,
\ee
which contributes to the integral in \rf{S2}
\be
\frac{K}{(1-\nu)^2} \int\d\theta_1 \int \d\theta_2 
\left(a^2 \theta_1 \theta_2+b^2\right) 
\frac{\dot\beta^2 (\theta_1)(\theta_2-\theta_1)^2}{
(\theta_2-\theta_1)^2 \ln^2 \frac{8}{1-\nu}}
\propto  \frac{1}{(1-\nu)^2}
\ee 
modulo the powers of $b/a \propto \left( \ln \frac8{(1-\nu)}  \right)^{-1}$.
The same contribution comes also from 
the domain of both  $\theta_1$ and $\theta_2$ near $\pi$. We thus have
\be
A_2 \propto \frac{1}{(1-\nu)^2} 
\ee
for every mode. 

Integrating the Gaussian integral for every mode and
using the $\zeta$-function regularization, we get
(disregarding the logs to this order) a pre-factor of the type
\be
\prod_{\rm modes} {(1-\nu)}\propto \left(\frac1
 {\sqrt{1-\nu}}\right)^4 \propto \exp{\left(\frac{\pi^2}{2}\frac{a}b \right)},
\label{Lue}
\ee
where the product runs over 4 sets of modes, which results in
a factor of 4 in the exponent.
This coincides with the L\"uscher term~\rf{lut} for a
rectangle of the size $R\times T$ in $d=26$ dimensions 
provided%
\footnote{It is worth noting that the Wilson loops for a rectangle
and ellipse then coincide~\cite{ABG81} in $d=4$ (the only dimension with a 
$T/R$ Coulomb term) to the second order 
of perturbation theory.}
\be
\frac T R =\frac {\pi a}{2b}.
\label{provided}
\ee
In Appendices \ref{appC} and \ref{appD} we confirm this by
an explicit calculation of the determinant of the Laplace operator
and the conformal anomaly for an outstretched ellipse.

\section{A generalization to arbitrary dimensions\label{s:5}}

The results of two previous sections demonstrate the already
mentioned fact that the ansatz \rf{ansatz} has to be modified in order to
describe the L\"uscher term in $d=4$ dimensions.

A simple modification is based on the form
of the path integral for a rectangle 
\be
\int \D\beta(t) \e^{-A_2[\beta(t)]}
\stackrel{T\gg R}\propto \e^{\pi T/R} 
\label{III2}
\ee
with quadratic action $A_2[\beta(t)]$ given by \eq{r-aA2}. 

For an arbitrary curve this path integral can be expressed through 
the determinant of the corresponding operator, that enters $A_2$,
which we denote as
${O}$:
\be
\int \D\beta(t) \e^{-A_2[\beta(t)]} =
\left( \det {O} \right)^{-1/2}. 
\label{OO2}
\ee
It is now clear that the following modification of
the ansatz~\rf{ansatz} will provide the correct value of the 
L\"uscher term~\rf{lut} in $d$ dimensions: 
\begin{equation}
\int {\cal D}t \e^{-KA[x(t)]}~(\det O)^{-(d-26)/48}.
\label{pp8}
\end{equation} 
This modification of Eq.~(\ref{ansatz})  does
not effect the classical limit, leading to the area law, and has the meaning 
of altering a pre-exponential in the semiclassical approximation.

To make the structure of the operator ${O}$ more explicit, it is convenient
to use the expansion \rf{expa} of the direct function
$\theta(\sigma)$ rather than that \rf{expa1} of the inverse function 
$\sigma(\theta)$ as above. 
Using the identity
\be
\beta(t_*(s))=-\frac{1}{\frac{\d t_*(s)}{\d s}} \,\beta(s)\,, 
\ee
which stems from the definitions~\rf{expa} and \rf{expa1},
we then obtain for the real-axis parametrization:
\be
A_2\left[\beta(s)\right] = \frac K{4\pi}  \int\d s_1 \pint \d s_2 
\frac{\dot x(t_*(s_1))\cdot \dot x(t_*(s_2))}{(s_1-s_2)^2 } 
\left[\beta(s_1)-\beta (s_2)\right]^2,
\label{AAA2}
\ee
which determines the ``momentum'' (with respect to $s$) space operator
\be
O(p_1,p_2)= \frac{\K}{8\pi} \int \d q |q| 
\Big(
2  \dot x(p_1+q)\cdot  \dot x(-p_2-q)-\dot x(p_1-p_2+q) \cdot \dot x(-q) 
- \dot x(q)\cdot  \dot x(p_1-p_2-q)\Big)
\ee
with 
\be
\dot x (p) \equiv \int \d s \e^{\i p s} \dot x(t_*(s))\,.
\ee
We can finally substitute $t_*(s)$ by $t(s)$ in this formula without
changing the semiclassical approximation.

It is worth noting that in contrast to the Laplace operator 
of Ref.~\cite{DOP84}, 
where the L\"uscher term was obtained for the Polyakov string, the present
operator $O$ lives in the boundary, which makes the construction nontrivial.

\section{Conclusions\label{s:6}}

The conclusion is that the reparametrization of the boundary curve involved in 
Eq. (\ref{ansatz}) carries information on the transverse fluctuations in 26
dimensions. As is shown in the previous section, 
it is possible to generalize this to any dimensions.

Our motivation for the present paper is our previous work on
the QCD/string scattering amplitudes \cite{MO08}, where we used that
the amplitude can be expressed in terms of a Wilson loop through 
Feynman path integration. There we only
considered the leading area behavior. However, having developed a 
path integral expression for the next term, we hope that the $x^\mu$
integrals can be performed, thereby providing a momentum space
analogue of the L\"uscher term.
We hope this may help to answer a very interesting question as to
how the intercept of the Regge trajectory changes under such
a modification of the ansatz for the Wilson loops. 

\begin{acknowledgments}
We are indebted to Andrey Mironov and Niels Obers  
for useful discussions.
\end{acknowledgments}

\appendix

\section{Measure for integrating over reparametrization\label{appA}}

Introducing
\be
v_i=s_i-s_{i-1} \qquad s_N=s_f\,,
\label{127}
\ee
we rewrite the measure of \cite{MO09} for the integration over 
reparametrizations as
\bea
\int_{s_0}^{s_f}\D_{\rm diff} s &\equiv &\lim_{N \to \infty}
\prod_{i=2}^{N-1}\int_{s_0}^{s_{i+1}}
 \frac{\d s_i}{(s_{i+1}-s_{i})}\int_{s_0}^{s_{2}}
 \frac{\d s_1}{(s_{2}-s_{1})(s_{1}-s_{0})} \non &=&
\lim_{N \to \infty}\prod_{i=1}^{N} \int_{0}^{\infty} \frac{\d v_i}{v_i}\, 
\delta^{(1)}\Big( s_f-s_0-\sum_{j=1}^{N} v_j \Big).
\label{vvv}
\eea

The integration over $v_i$'s in \eq{vvv} can be represented through 
the integration over a scalar field as follows.
Writing
\be
v_i=\e^{\psi_i/2}\,,
\label{127p}
\ee
 we have
\be
\int_0^\infty \frac{\d v_i}{v_i} \cdots = \frac 12
\int_{-\infty}^{+\infty} \d \psi_i 
\ee
and
\be
\int_{s_0}^{s_f}\D_{\rm diff} s =
\lim_{N \to \infty}\frac{1}{2^{N}}
\prod_{i=1}^{N} \int_{-\infty}^{+\infty} {\d \psi_i}\, 
\delta^{(1)}\Big( s_f-s_0-\sum_{j=1}^{N} \e^{-\psi_j/2} \Big).
\label{vvvi}
\ee

This represents the continuous measure as
\be
\int_{s_0}^{s_f}\D_{\rm diff} s = \int \D \psi \, 
\delta^{(1)} \Big(s_f-s_0-\int_{s_0}^{s_f} \d t \e^{-\psi(t)/2} \Big),
\label{continual}
\ee
where $t$ is a certain parametrization of the contour (e.g.\ through the
proper time) and $\D \psi$ is the usual measure
\be
\int \D \psi = \prod_{s=s_0}^{s_f}\int_{-\infty}^{+\infty} \d\psi(s) \,.
\ee

The scalar field $\psi$, that appears in Eqs.~\rf{127}, \rf{127p},
is in fact a discretization of the boundary value of the Liouville field 
\be
\varphi(\tau,s)\Big|_{\rm boundary}= \psi(s)\,,
\label{defpsi}
\ee 
which is related to the boundary metric:
\be
\sqrt{\left(\frac{\d x^\mu(s)}{\d s}\right)^2}
=\e^{\psi(s)/2}\,.
\label{112}
\ee
Under the reparametrization $s\to f(s)$ ($\d f/\d s\geq 0$), 
the boundary value of the Liouville field $\psi(s)$ transforms as
\be
\psi(s) \to \psi(f(s))=\psi(s)-2 \ln \frac{\d f(s)}{\d s}\,,
\label{115}
\ee
that clarifies its relation to reparametrizations.

The results of this Appendix make it possible to relate 
the ansatz~\rf{ansatz} with Eq.~(17) of Ref.~\cite{DOP84},
where the path integral over $\psi$ is the same as the
path integral over reparametrization in view of Eq.~(11) of \cite{DOP84}.
They coincide in $d=26$ provided $S_{\rm cl}$ is the Douglas integral. 

\section{Douglas' algorithm for plane contours: conformal map\label{appB}}

The construction of the coordinates of the minimal surface, enclosed by 
a {\em plane}\/ contour, is given by conformal mappings.
Let us describe such a contour by two functions
$x_1(t)$ and $x_2(t)$.
Motivated by Appendix~H of Ref.~\cite{Mig94}, we define the 
analytic functions
\be
\Phi_\mu(z)=\int_{-\infty}^{+\infty} \frac{\d s'}{\pi}\,
\frac{\left[x_\mu(t_*(s'))-x_\mu(t_*(z))\right]}{(s'-z)^2}.
\label{Phi}
\ee

The real and imaginary parts of $\Phi_\mu$ are 
\begin{subequations}
\bea
\Re \Phi_\mu(s)&=&  \pintii \frac{\d s'}{s'-s}\, \dot x_\mu(t_*(s')), 
\label{RI1}\\
\Im \Phi_\mu(s)&=&  \frac{ \d x_\mu(t_*(s))}{\d s} = \dot x_\mu(t_*(s)) 
\frac{\d t_*(s)}{\d s}.
\label{RI2}
\eea
\label{RI}
\end{subequations}
Therefore, Douglas' minimization equation 
\be
\pintii \d s' \,\frac{\dot x(t_*(s))\cdot \left[ 
 x(t_*(s))- x(t_*(s'))\right]}{(s-s')^2} =0.
\label{Dou1}
\ee 
is satisfied if 
\be
\sum_\mu \Im \Phi_\mu^2(s)=0
\label{ifim}
\ee
at the real axis.   

For the circle we have 
\be
x_1(t)+\i x_2(t) \equiv C(t)=\frac{\i-t}{\i+t}
\label{circl}
\ee
so \eq{Phi} gives
\be
\Phi_1(z)=\i \Phi_2(z)=\frac{2}{(\i+z)^2}
\label{circlPhi}
\ee
with
\be
\sum_\mu \Phi^2_\mu(s) =0, 
\label{obe}
\ee
which obviously obeys \eq{ifim}.
>From Eqs.~\rf{circlPhi} and \rf{circl} we find
\begin{subequations}
\bea
 \Im \Phi_1(s) = -\frac{2s}{(1+s^2)^2} ,&& \qquad \frac 12 \dot x_1(t)=
- \frac{2t}{(1+t^2)^2}; \\* 
  \Im \Phi_2(s)=\frac{1-s^2}{(1+s^2)^2} ,&& \qquad \frac 12 \dot x_2(t)=
 \frac{1-t^2}{(1+t^2)^2} 
\eea
\end{subequations}
and from \eq{RI2} conclude that $t_*(s)=s$, as it should be for the circle.

For the functions obeying \eq{obe}, we always have $\Phi_1(s)=\i \Phi_2(s)$
and
\begin{subequations}
\bea
\Phi(s)\equiv \Phi_1(s)+\i\Phi_2(s) = 2\Phi_1(s);\\*
\Im \Phi_1(s)+\i \Im \Phi_2(s)=-\frac{\i}{2} \Phi(s).
\eea
\end{subequations}
Equation~\rf{RI2} can then be rewritten as
\be
-\frac{\i}2 \Phi(s)=\frac{ \d C(t_*(s))}{\d s} = \dot C(t_*(s)) 
\frac{\d t_*(s)}{\d s}.
\label{newPhi}
\ee
For the given analytic function $C(z)$ it determines the 
reparametrizing function $t_*(s)$.

For the circle and ellipse it is more convenient to use the unit-disk 
parametrization, when 
\be
\Phi(\omega) = (\omega+1)^2 \oint\nolimits_{S^1} \frac{\d\omega}{2\pi\i}
\frac{\left[C(\omega)-C(\omega')  \right]}{(\omega-\omega')^2}.
\label{Phi(om)}
\ee
For the ellipse $C(\omega)$ is given by the conformal map (where  $\nu$
is the same as $s$ in Eq.~(B26) of \cite{MO09})
\be
C(\omega)=\sqrt{a^2-b^2} \sin \left[ \frac{\pi}{2K(\nu)} \,
F\left(\frac{\omega}{\sqrt{\nu}},\nu \right) \right]
\ee
and from \eq{Phi(om)} we obtain
\bea
\Phi(\omega)&=&\sqrt{a^2-b^2} 
\frac{\pi}{2K(\nu)} \,
\frac{(\omega+1)^2 }{\sqrt{\nu-\omega^2}\sqrt{1-\nu \omega^2}}
\cos \left[ \frac{\pi}{2K(\nu)} \,
F\left(\frac{\omega}{\sqrt{\nu}},\nu \right) \right]\non
&=&(\omega^2+1) \frac{\d C(\omega)}{\d \omega}.
\label{ellipsePhi}
\eea

The final step is to extract $\d \theta_*(\sigma)/\d \sigma$ from
\eq{newPhi}. Remembering that 
\be
C(\theta)= a \cos \theta + \i b \sin \theta,\
\ee
we find
\be
\frac{\d \theta }{\d \omega}= 
\frac{1}{\sqrt{\nu-\omega^2}\sqrt{1-\nu \omega^2}}
\ee
which for $\omega=\e^{\i\sigma}$ reproduces Eq.~(B32) of \cite{MO09}.

\section{Calculating the L\"uscher term for an ellipse\label{appC}}

\subsection{Elliptical coordinates and Mathieu functions}

Let us consider an ellipse as the boundary contour.
To calculate the determinant of the Laplace operator with this boundary
conditions, we parametrize the surface by elliptical (Lam\'e) coordinates
\be
x_1= h\, \cosh \tau \cos \sigma\,, 
\qquad x_2=h\, \sinh \tau \sin \sigma 
\ee
or
\be
x_1+\i x_2= h \cosh\left(\tau+\i\sigma \right).
\label{C2}
\ee
Here $\sigma \in [0,2\pi)$ and $\tau \in [0,\tau_0]$.
The boundary is approached for $\tau=\tau_0={\rm arctanh}\; b/a$ and
foci are at $(\pm h,0)$ with
\be
h=\sqrt{a^2-b^2}.
\ee

The elliptic coordinates are conformal:
\be
\d s^2=h^2 \left(\sinh^2 \tau+\sin^2 \sigma \right)
\left(\d \tau^2+\d\sigma^2 \right),
\ee 
so the area in the conformal gauge 
reads
\be
A=\frac 12 \int _0^{\tau_0} \d \tau \int_0^{2\pi} \d \sigma \, 
\left( \frac{\partial X}{\partial \tau}\cdot 
\frac{\partial X}{\partial \tau}+  \frac{\partial X}{\partial \sigma}\cdot 
\frac{\partial X}{\partial \sigma}   \right) 
\ee
The Laplacian reads in elliptic coordinates as
\be
\Delta = \frac{1}{h^2 \left(\sinh^2\tau+\sin^2 \sigma \right)}
\left( \frac{\partial^2}{\partial \tau^2}+  
\frac{\partial^2}{\partial \sigma^2}  \right).
\ee

The Laplace equation separates in elliptic coordinates to the Mathieu 
equations
\begin{subequations}
\bea
\frac {\d ^2 F}{\d \sigma^2}+\left(\alpha -2q \cos 2\sigma\right)F&=&0\,, \\
\frac {\d ^2 G}{\d \tau^2}-\left(\alpha -2q \cosh 2\tau\right)G&=&0\,. 
\eea
\end{subequations}
where $\alpha$ is a separation constant and $q$ is related to the
eigenvalue $\lambda$ by
\be
q=\lambda \frac{h^2}4\,.
\ee
A complete set of solutions for $F$ and $G$ is given, respectively,  
by the Mathieu functions $ce_m(\sigma,q)$, $se_m(\sigma,q)$ and
the modified Mathieu functions $Ce_m(\tau,q)$, $Se_m(\tau,q)$ 
of integral order $m$. 

While the characteristic numbers $\alpha_m(q)$ and $\beta_m(q)$ for
the $ce_m$ and $se_m$ modes are not explicitly known for large $q\sim m^2$, 
which would be the case for an outstretched ellipse with $b/a\ll1$, 
 two asymptotic formulas exist~\cite{McL47}
\begin{subequations}
\bea
\alpha_m=\beta_m&\stackrel{m^2\gg q}=& m^2+\frac12 \frac{q^2}{m^2}+
\frac{5}{32}\frac{q^4}{m^6}+\frac{9}{64}\frac{q^6}{m^{10}}+
\frac{1469}{8192}\frac{q^8}{m^{14}}+\frac{4471}{16384}\frac{q^{10}}{m^{18}}+
\ldots \label{exp1}\\
\alpha_m=\beta_m&\stackrel{m^2\ll q}=&-2q+4m q^{1/2}-\frac12 m^2-
\frac1{2^4}\frac {m^3}{q^{1/2}} -\frac 5{2^8}\frac{m^4}q-
\frac{33}{2^{12}}\frac{m^5}{q^{3/2}} -\frac{63}{2^{14}}\frac{m^6}{q^2}-
\frac{527}{2^{18}}\frac{m^{7}}{q^{5/2}}-\ldots  \!\!\!\!\!
\non &&\label{exp2}
\eea
\end{subequations}
These expansions are not formally applicable for $c=q/m^2\sim 1$,%
\footnote{Some results concerning the Mathieu characteristic numbers 
for large $q \sim m^2$ can be found in Ref.~\cite{Alh96}.}
but the first formula works numerically for $c \la 0.7$, while
 second formula works numerically for $c\ga 0.4-0.5$, so there is
an overlap. Special values of $c$ are $c=0.61$ 
and $c=0.64$, when
\be
\alpha_m(cm^2)=\beta_m(cm^2)=2c m^2=2q\,.
\ee
They are obtained numerically with Mathematica, but are
derivable from the expansions \rf{exp1} and \rf{exp2}, correspondingly.
As we see from Fig.~\ref{fi:yield},
\begin{figure}
\vspace*{3mm}
\includegraphics[width=8cm]{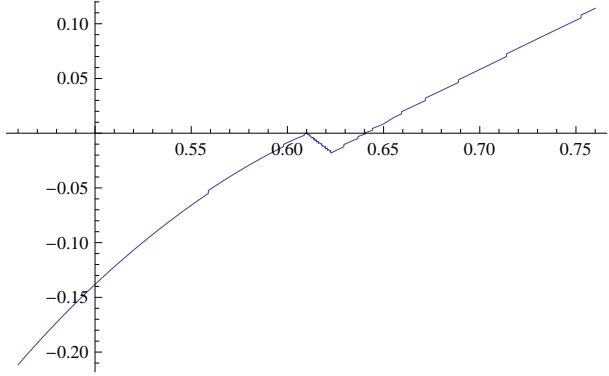}   
\caption[]{Plot of $1-\alpha_m(q)/2q$ versus $c=q/m^2$.}   
\label{fi:yield}
\end{figure}
the difference $2q-\alpha_m(q)$ is positive for $c>0.64$, where
the expansion \rf{exp2} is applicable.

The boundary condition requires $Ce_m(\tau_0)=0$ and $Se_m(\tau_0)=0$,
which  implies
\be
\tau_0= Z_{m,n}\,,
\ee
where $Z_{m,n}$ are the corresponding zeros. This determines the
eigenvalues of the Laplacian with $m$ and $n$ playing the role of 
angular and radial quantum numbers, respectively.
For small $\tau_0$ (= large $a/b$) we have~\cite{McL47}
\begin{subequations}
\bea
Ce_m(\tau)&\approx & 
{\rm const.}\times\cos \left(\sqrt{2q-\alpha_m}\tau\right) \\
Se_m(\tau)&\approx & {\rm const.}\times\sin \left(\sqrt{2q-\beta_m}\tau\right)  
\eea
\end{subequations}
so that
\be
\sqrt{2q-\alpha_m}\tau_0=\frac{\pi n}{2}
\ee  
because $\alpha_m=\beta_m$ for large $m$. Here $n$ is odd or even
for the Ce or Se modes, respectively.
We conclude therefore that
\be
2q-\alpha_m=\left(\frac{\pi n a}{2b}\right)^2
\ee
has to be large and justify large $q\sim m^2$.

\subsection{Evaluating the determinant}

Each mode contributes
\be
\det \left(\Delta^{-1/2}  \right)=\prod_{n,m} \lambda_{n,m}^{-1/2} 
=\e^{\sum_{n,m}\ln\left(\lambda_{n,m}^{-1/2}\right)}\,.
\ee
Since $a$ is large, the sum over $m$ can be replaced by an integral over 
$\omega=m/a$ like in Ref.~\cite{Alv81} and for large $q^{1/2}\sim m \sim a/b$, 
we have
\be
\lambda_{n,m}^{1/2}=r f\left(\frac {\omega}{r }\right),
\ee
where 
\be
r=\frac{\pi n}{2b}\,,
\ee
so that
\be
\sum_{m}\ln\left(\lambda_{n,m}^{-1/2}\right) =-
a \int_0^\infty \d \omega\, \ln \left[r f\left(\frac {\omega}{r }
\right)\right].
\ee
For large $r$
the integral on the right-hand side is proportional to $r$ and
we can get the coefficient of proportionality by differentiating
with respect to $r$. This gives
\be
\sum_{m}\ln\left(\lambda_{n,m}^{-1/2}\right) =-
a r \int_0^\infty \d x\, \left[1- x f'(x)/f(x)\right]
\ee
and finally we obtain
\bea
\sum_{n,m}\ln\left(\lambda_{n,m}^{-1/2}\right)&=&
-\sum_{n} \frac{\pi n a}{2b}\int_0^\infty \d x\, \left[1- x f'(x)/f(x)\right]
\non &=&\frac{\pi}{24}\frac a b 
\int_0^\infty \d x\, \left[1- x f'(x)/f(x)\right].
\label{alva}
\eea

For an $T\times R$ rectangle with $T\gg R$, when
\be
\lambda_{n,m}=\frac{\pi^2 n^2}{R^2} +\frac{\pi^2 m^2}{T^2} 
\qquad \fbox{rectangle}
\ee
resulting in 
\be
f(x)=\sqrt{1+x^2}\qquad \fbox{rectangle}\,,
\ee
we reproduce the well-known result for the L\"uscher term
by the substitution $a=T/\pi$, $b=R/2$.

For an outstretched ellipse we have from the expansion~\rf{exp2}
\be
f(x)\stackrel{x<x_0}=1+x+\frac{x^2}{4}-\frac{x^3}{16}-
\frac{x^4}{128}+\frac{15 x^5}{1024}
-\frac{9 x^6}{2048} 
+\ldots
\label{ma1}
\ee
for $x<x_0\approx 1.95$, where the series converges, and
\bea
f(x)&\stackrel{x>x_0}=&1.600095 x + \frac{0.960520}x - 
 \frac{1.041899 }{x^3} + \frac{2.418227 }{x^5} - 
 \frac{7.138640}{x^7} + \frac{23.73152 }{x^9} \non &&- 
 \frac{84.70585 }{x^{11}} + \ldots 
\label{ma2}
\eea
for $x>x_0 \approx 1.95$.
They match each other pretty well, as is depicted in Fig.~\ref{fi:match}.
\begin{figure}
\vspace*{3mm}
\includegraphics[width=8cm]{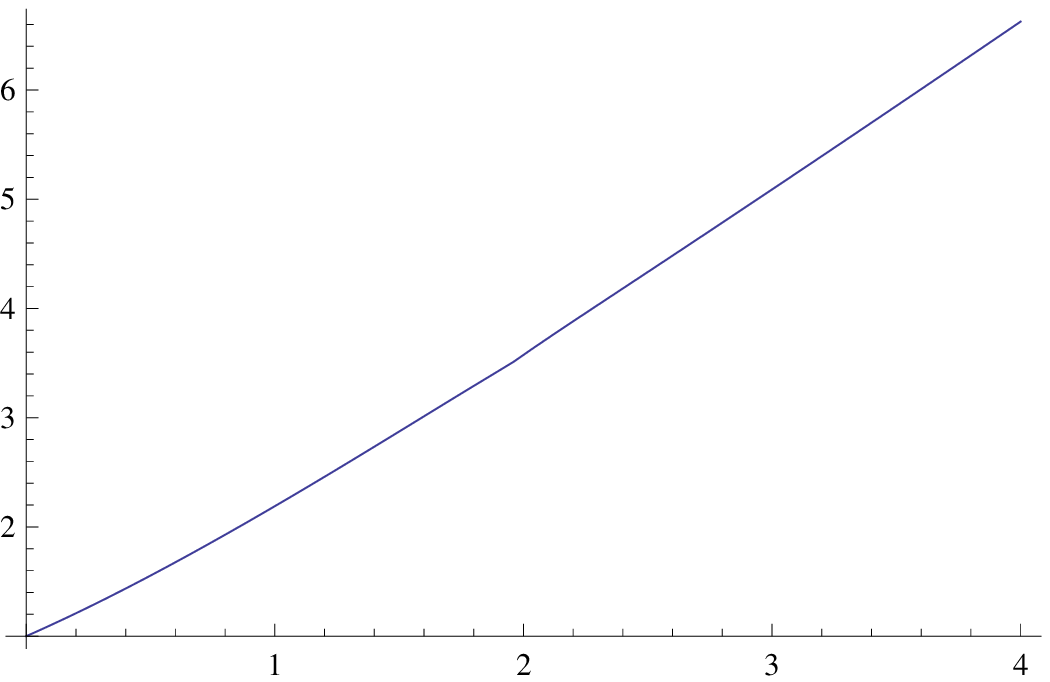} 
\includegraphics[width=8cm]{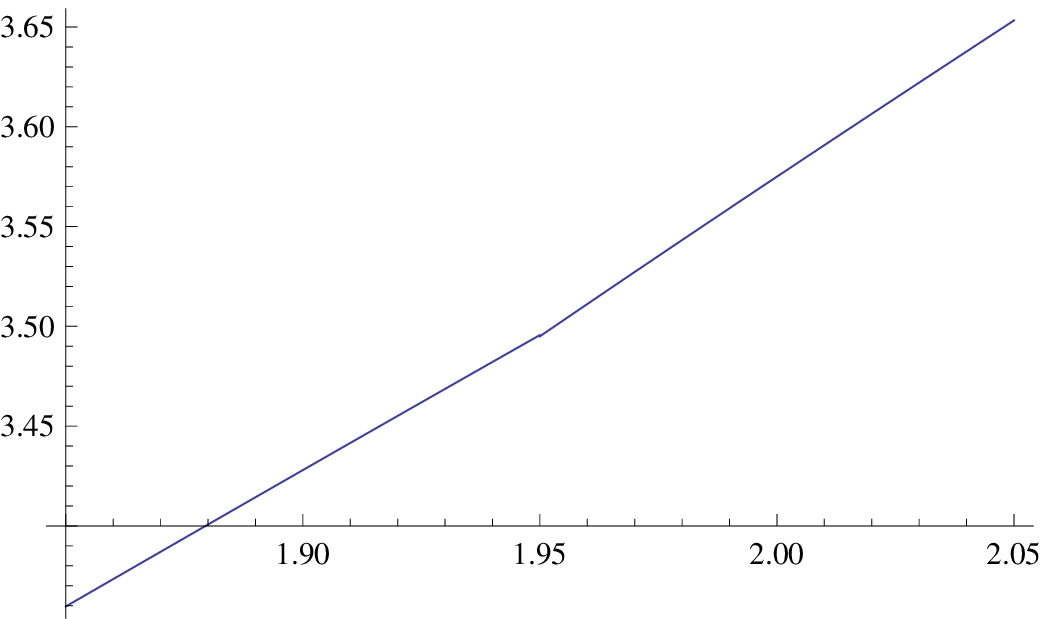}   
\caption[]{Matching the expansions \rf{ma1} and \rf{ma2}.
The region near $x=1.95$ is enlarged in the right figure.}   
\label{fi:match}
\end{figure}
The numerical value of the integral on the right-hand side of \eq{alva} is
then 2.84 to be compared with $\pi$ as is expected from \eq{Lue}.

To clarify the calculation, let us keep only three terms on the right-hand 
side 
of \eq{exp2}, when we get a quadratic equation and analytical formulas
are available.
We then find 
\be
f(x)=x+ \sqrt{1+\left.x^2\right/2}
\ee
and expanding it either in $x$ or in $1/x$ we obtain analogs
of Eqs.~\rf{ma1} and \rf{ma2}. The two expansions now match at
$x_0=\sqrt{2}=1.41$ to be compared with $x_0=1.95$ for all nine terms
left on the right-hand side of \eq{exp2}. The value of the integral 
is now $\sqrt{2} \log\left[3+2 \sqrt{2}\right]=2.49$ which is
smaller than $2.84$, which may characterize to which extend
the approximation of $\alpha_m$ by the eight term in \rf{exp2}
is better than by the tree terms. In Appendix~\ref{appD} 
we confirm the value of $\pi$ by another method.   

\section{L\"uscher term from conformal anomaly\label{appD}}

A starting point is the formula~\cite{luscher,Aff86} 
\be
\hbox{L\"uscher term}\;=\frac{1}{96\pi} \int \d \tau \d\sigma\,
 \partial_a \ln \rho  \,\partial_a \ln \rho 
\ee
that relates the L\"uscher term to the conformal anomaly. 
Here the metric $\rho$ is
\be
\rho = \left| \frac{\d z}{\d u}\right|^2 \,, \qquad u=\tau+\i \sigma
\label{rhometric}
\ee
where 
\be
\omega= h \cosh \left(\tau+\i\sigma \right)
\ee
runs inside the ellipse with $\tau$ and $\sigma$ being elliptic coordinates 
as in \eq{C2}. The function (inverse to (B23) from \cite{MO09})
\be
z(\omega)= \sqrt{\nu} \sn \left(\frac{2 K}{\pi}
\left(\frac{\pi}2-\sigma+\i \tau \right)\right)
= \sqrt{\nu} \sn \left(K-\frac{2 K}{\pi}
\left(\sigma-\i \tau \right)\right)
=\sqrt{\nu} 
\frac{\cn \left(\frac{2 K}{\pi}
\left(\sigma-\i \tau \right)\right)}{\dn \left(\frac{2 K}{\pi}
\left(\sigma-\i \tau \right)\right)}
\label{ellidisk}
\ee
conformally maps ellipse onto a unit disk.

To calculate \rf{rhometric} for $a/b\gg1$, we substitute
\be
\frac{\cn\left( \frac{2 K}{\pi}\i u,\nu\right)}
{\dn \left( \frac{2 K}{\pi}\i u,\nu\right)}=
\frac{1}{ \dn\left(\frac{2 K}{\pi}u,\nu'\right)}
\ee
with
\be
\nu'=\sqrt{1-\nu^2}\to 0 \,.
\ee
Differentiating, we get
\be
 \frac{\d z}{\d u} = \sqrt{\nu}\frac{2 K}{\pi} (1-\nu^2)
\frac{\sn\left(\frac{2 K}{\pi}u,\nu'\right)
\cn\left(\frac{2 K}{\pi}u,\nu'\right)}
{\dn^2\left(\frac{2 K}{\pi}u,\nu'\right)}\,.\label{dddd}
\ee
Using the fact that Jacobi elliptic functions reduce to trigonometric functions
as $\nu'\to0$ and substituting
\be
K\stackrel{\nu\to1}\to \frac{\pi^2}{8}\frac ab\,,
\ee
we infer from \eq{dddd}
\be
 \frac{\d z}{\d u} \propto \exp{\left(-\i\frac\pi 2 \frac a b u\right)} \quad
\Longrightarrow \quad \rho \propto \exp\left(\pi \frac ab \sigma \right)
\ee
and
\be
\hbox{L\"uscher term}\;=\frac{1}{96\pi} \int_0^{\tau_0\approx \frac ba} 
\d \tau \int_0^{2\pi}\d\sigma\,
 \left(\pi  \frac a b \right)^2 =\frac{\pi^2}{48} \frac ab 
\ee
that confirms the extra $\pi/2$ in \eq{Lue} for an ellipse in comparison
with a rectangle.


\end{document}